# Vowels and Prosody Contribution in Neural Network Based Voice Conversion Algorithm with Noisy Training Data

Olaide Ayodeji Agbolade

*Abstract*—**This research presents a neural network based voice conversion (VC) model. While it is a known fact that voiced sounds and prosody are the most important component of the voice conversion framework, what is not known is their objective contributions particularly in a noisy and uncontrolled environment. This model uses a 2-layer feedforward neural network to map the Linear prediction analysis coefficients of a source speaker to the acoustic vector space of the target speaker with a view to objectively determine the contributions of the voiced, unvoiced and supra-segmental components of sounds to the voice conversion model. Results showed that vowels "a", "i", "o" have the most significant contribution in the conversion success. The voiceless sounds were also found to be most affected by the noisy training data. An average noise level of 40 dB above the noise floor were found to degrade the voice conversion success by 55.14 percent relative to the voiced sounds. The result also shows that for cross-gender voice conversion, prosody conversion is more significant in scenarios where a female is the target speaker.**

*Index Terms*—**Formant, Prosody, Voice Conversion, Neural Network**

## I. INTRODUCTION

Voice conversion refers to a well-established signal processing technique used to make words uttered by one person usually the source speaker sound like another person, the target speaker [1]. This signal processing technique is a well-researched area that has experienced lots of improvement from decades of research efforts at solving challenges associated with the process. Application areas for this signal processing technique include text to speech synthesis, voice-over, gamming applications, identity modeling for security threat analysis among several others [2]. Two popular research paths have been taken by researchers of all ages. The first been concatenative synthesis [3] while the other is parametric [4]. It may be argued if the conventional concatenative approach can be fully regarded as a voice conversion technique since in the real sense of things, nothing is converted. The processes involve breaking down speech samples from a speaker into phonemes which are processed before being stored in a database. To synthesis new speeches, the phonemes are combined discretely and synthesized. In this method, there is no source or target speaker, only speeches. While it is true that the speech produced by this method sounds very natural particularly if the problem of discontinuities at the concatenation points can be removed [5], they are however still cumbersome and cannot be readily extended to new speakers [6]. Parametric voice conversion on the other hand can easily morph several source speakers into a single target speaker whose voice model has been built beforehand.

In parametric voice conversion, the two components usually involved are the spectral component of the speaker and the prosody [7]. The spectral structure of the speech largely depends on the formants which are built from the vowels and other voiced component of the speech. The prosody is the supra-segmental part of the speech that include things like stress and intonation. They are also usually a good reflection of the speaker's anatomical make up.

The focus of this work is the objective contribution of these voiced sounds and prosody in a Neural Network based voice conversion framework. Particular attention was made to ensure the naturalness of the conversion process by ensuring the training data were not recorded in a noise proof condition like most voice conversion models do. Rather, speech data were obtained in an uncontrolled outdoor environment. The vowels in speech samples constitutes the bulk of the formants since they represent the voiced part of the speech and are thus the most important ingredient in the voice conversion process. The goal in this work is not to show all the voiced and unvoiced sounds but to reveal how the pronunciation of some of these sounds affect the voice conversion process. Linear Predictive coding was used for speech modeling while feedforward Neural Network (NN) was employed to map the spectral components of the source speakers to the acoustic vector space of the target speakers.

Our contributions are as follows. Firstly, in most voice conversion models, carefully selected training samples are used for model training. The training data are recorded in controlled environment with minimal noise interference. While this may improve performance, it results in a model that may not always guarantee reliable results when deployed in real life scenarios which is always noisy and uncontrolled. This work thus set out to provide submissions that are comparable to real life expectations. Secondly, it is a known fact that voiced sounds are of utmost important in the speech conversion process. What is however not obvious to the best of our knowledge is the objective contribution of these vowels in training samples recorded in uncontrolled environment. Lastly, the work reveals the effect of prosody in cross gender voice conversion.

## II. Related Works

Parametric voice conversion methods are one of the most reliable in term of flexibility. Reference [8] introduced one of the foremost voice conversion concept with the use of codebook mapping technique. He used vector quantization and spectrum mapping to create speaker's codebooks which contained approximate correspondence between different speakers' acoustic vectors obtained from both the source and target speakers. Codebooks for spectrum parameters, power values and pitch frequencies were used as weighting function for the conversion with experimental results showing that the converted speech has the characteristics of the target speaker.

Subsequent to the success reported by [8], further research efforts were made to improve upon prosodic and spectral features of both the source and target speakers. Reference [9] developed a voice cloning algorithm by using Dynamic frequency warping and linear multivariate regression for spectral features. The work used Pitch synchronous overlap add (PSOLA) [10] for prosodic transformation. The work found PSOLA as an effective tool for converting prosody of speeches. PSOLA works on the fundamental principle that if a signal is compressed (signal decimation), the pitch increases and when it is extended (signal interpolation), the pitch reduces. However, when compressed or extended, the timing changes and so to compensate for the change in the duration of the signal, it is resampled by the inverse of the proportion by which its duration was adjusted. Reference [11] also used time domain PSOLA algorithm to convert the prosody of a source speaker to that of the target by simply stretching or compressing the source speaker's speech sample in accordance with that of the target.

There are equally several neural network based voice conversion models in literature. For instance, [12] used artificial neural network to map the spectral features of a source to that of the target and found the ANN model to outperform Gaussian mixture models (GMM) in quality and intelligibility. Reference [13] used auto associative neural network to map the linear prediction coefficients of the source into the target acoustic space. The converted speech was synthesized with the linear prediction residual from the target and the LPC coefficients of the source. Mean opinion score and t-test were used to evaluate the performance of the system with the transformed speech adjudged to have the properties of the target. Reference [14] used a speaker independent deep auto encoder to make generalization of Mel-Cepstral coefficients of speech features. ANN was used to map the deep auto encoder for two sentences and seventy sentences training set. The deep neural network was found to out-perform the popular joint density Gaussian mixture model particularly when just two training sentences were used.

Reference [15] proposed a voice conversion algorithm using generative trained deep neural networks with multiple frame spectral envelopes. A four-layer deep neural network (DNN) performed the non-linear mapping of the spectral features of the source to the target. The DNN training was done layer by layer with two restricted Boltzmann machines and a bidirectional associative memory network. The work like most others used a well-constructed Chinese speech database and hence obtained a result that was comparable to most neural network based voice conversion model. An improvement to the model developed in [15] was proposed in [16]. The improved model used DNN with generative layer-wise training to mitigate the problem of spectral loss and inadequacy of joint density Gaussian Mixture Model (JDGMM) spectral features distribution. An improvement in naturalness of the converted speech was observed. A voice conversion model which adopts sequence error minimization training of neural networks was proposed in [17]. The model aimed at overcoming the limitations associated with neural network voice conversion model trained on frame error minimization criterion by proposing a model, based on sentence optimization and minimum generation error training on HMM based synthesis. Significant improvement on the naturalness of the converted speech were also reported.

Reference [18] proposed a general regression neural network (GRNN) based voice conversion model. Both Line Spectral Frequencies (LSFs) and Linear Prediction residues were used to model the acoustic vector space of the source and target speaker. The study identified artifacts in consecutive speech frames as a major problem in the GRNN model. This challenge was however mitigated with the adoption of wavelet packet decomposition. By successfully mapping the LSFs, pitch contour and energy profile of the source speaker into vector space of the target, a slightly better performance than conventional voice conversion models were reported.

A Recurrent Temporal Restricted Boltzmann Machine based voice conversion model was proposed by [19] and [20]. The models were found to outperform the conventional Gaussian mixture model and neural network voice conversion method. A deep bidirectional long short-term memory based recurrent neural networks was proposed by [21] to improve the naturalness of voice conversion models. To overcome the need for parallel data in building conversion function for VC models [22] proposed a novel architecture that uses a cycle-consistent adversarial network.

For all the research efforts made in voice conversion, little efforts have been made to study objective contribution of formants and vowels in the speech conversion process. For instance, [23] worked on bilingual voice conversion using weighted frequency warping based on formant space. The work only used text to speech generated speech instead of real human speeches. Formant spaces were modeled by the monophthongs of the target language but results were only described as promising to say the best. Reference [24] used deep neural networks for voice conversion of synthesized speeches. The work used a deep neural network in a linear transform function of formants. The formants features were only used as additional information to the deep neural model conditioning but the objective contribution of these formants were never studied.

A more recent study by [25] on prosody also only revealed its impacts in Wh-question and declarative statements in Mandarin Wh-words. It is therefore necessary to study objective contributions of formants and voiced sounds in a neural network based voice conversion model.

## III. RESEARCH METHODOLOGY

The architecture for the proposed NN based voice conversion model is presented in Fig. 1. This framework has been extensively discussed in [2]. The voice conversion model development is in two stages. In the first stage, the training of the NN model was done while the second phase involves the testing of the model. In the training phase a database which consist of parallel utterances from four speakers was used. The speakers consist of two males and two females within the age bracket of 25 and 28 years. The average pitch of the speakers is shown in table I. One hundred phonetically balanced words were recorded and sampled at 11025 samples per second. These Phonetically balanced words are group of words presented in [26] that has the various English language phonemes occurring at approximately the same frequency at which they are used in every day to day conversation thus making them suitable for speech training. Each of the words spans duration of 0.62 second. The words are framed and windowed in order to prepare them for linear prediction analysis. The average noise level during recording is about 40 DB above the noise floor.

### A. Vocal Tract Modelling

The first part of the model involves the use of Linear Prediction Analysis to model the vocal tract of the speakers during the pronunciation of voiced sounds with particular attention to the vowels which are mapped out using phoneme segmentation techniques. The LPC filter used to achieve this was modeled as an all pole filter transfer function given by equation 1 where p is number of poles and $a_i$ the filter coefficients.

$$H(z) = \frac{1}{\sum_{i=0}^{p} a_i z^{-1}}$$ (1)

Prior to the LPC analysis, the training samples were pre-emphasized with a finite impulse response filter given by equation 2 to ensure the speech samples have energy level which is fairly constant across the entire speech duration.

$$H(Z) = 1 - 0.95 Z^{-1}$$ (2)

The speech samples were divided into frames of 25 milliseconds with a 20 milliseconds overlap to ensure quasi-stationarity of the speech signal. Discontinuities at the edges of the frames were taken care of with a Gaussian window function given by equation 3.

$$W(n) = e^{\frac{-(n-m)^2}{2(\sigma N)^2}} \qquad n = 0,1,2,..,N-1$$ (3)

Where $m$ equals $\frac{(N-1)}{2}$, $\sigma$ standard deviation and $N$ the length of the window.

Equation 4 is the autocorrelation function which was used to acquire energy concentration region in each of the signal frame

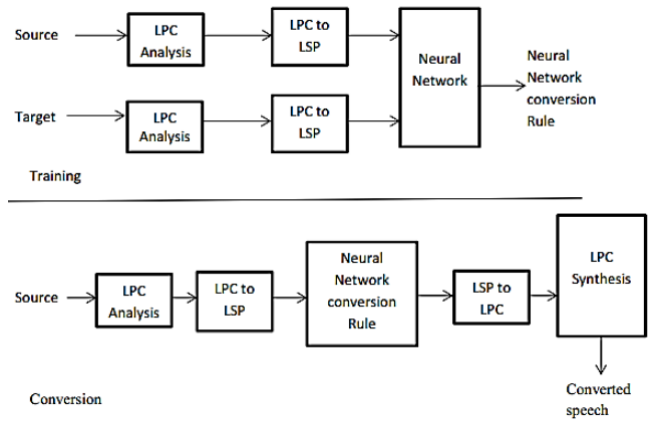

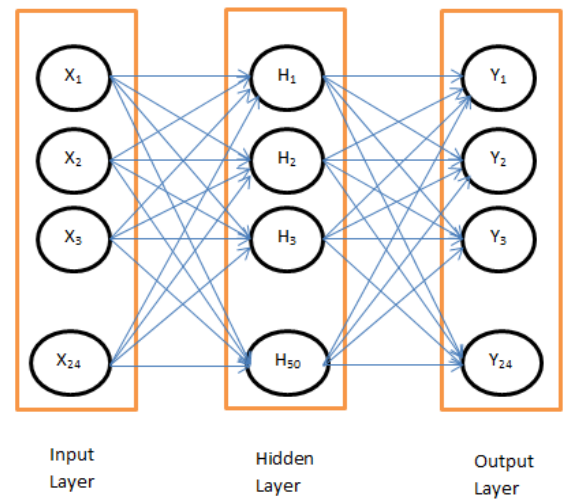

Fig. 1. Neural Network Voice Conversion Framework [2]

Fig. 2. 2-Layer Neural Network Architecture

TABLE I: PITCH INFORMATION OF SELECTED SPEAKERS [2]

| Speaker | Speakers' Pitch | | |
|---|---|---|---|
| | *Minimum pitch (Hz)* | *Minimum Pitch (Hz)* | *Average Pitch (Hz)* |
| Male 1 | 96.54 | 134.82 | 120.86 |
| Male 2 | 84.45 | 118.30 | 102.89 |
| Female 1 | 213.94 | 267.93 | 245.68 |
| Female 2 | 192.33 | 259.03 | 226.32 |

$$R_{xx}(\tau) = \lim_{N \to \infty} \frac{1}{N} \sum_{t=1}^{N} x(t) x(t+\tau)$$ (4)

Where $x(t)$ represents the windowed signal, N the total number of samples and $\tau$ the lag between subsequent frames

Maximum energy in the signal as shown in equation 5 is obtained at points where the lag between subsequent frames equals zero.

$$R_{xx}(0) = \lim_{N \to \infty} \frac{1}{N} \sum_{t=1}^{N} x^2(t)$$ (5)

The 24 order filter used in modelling the vocal tracts of source and target speakers in this work were obtained by using the Levinson-Durbin's algorithm to solve the resulting Toeplitz matrix from equation 5. The filters were derived using equation 6 [27].

$$p \geq 4 + \frac{f_s}{1000}$$ (6)

### B. Excitation Signal

The excitation signal contains the prosodic details of the speakers and are therefore very important in the speech conversion process as it reveals lots of details about the individuality of the speaker. It is calculated as the difference between the original signal and the linear prediction or simply as the error in the speech signal linear prediction. Equation 7 shows the process of obtaining the excitation signal of the speakers. It shows that by filtering the input signal with the inverse transfer function of the LPC filters, the excitation signal E (Z) can be obtained.

$$E(Z) = S(Z) \bullet \frac{1}{H(Z)} \qquad (7)$$

From the excitation signal obtained from both source and target, the prosody mapping process was completed by replacing the pitch of the source speaker with that of the target using the pitch synchronous overlap add method (PSOLA) [10].

### C. Neural Network Architecture

Neural network was used to define a function $f: x \rightarrow y$ which forms a generalized relationship between the acoustic vector space of source $x$ and target $y$. For training, the Levenberg-Marquardt feedforward backpropagation algorithm was used to find a function $f \in F$ iteratively until an optimal solution is obtained using a cost function $C = E[(f(x) - y^2)]$ for the dataset$(x, y)$.

The neural networks find the epoch that best represents the generalization of the two signals by iteratively calculating the mean square error of the signals. The neural network presents a training sample x, compute an output y for each input and compares the output y with target t while iteratively adjusting its weight and threshold using mean square error until all y equals t or an optimal solution that meets the error goal is obtained. A 24-50-24 neural network configuration shown in Fig. 2 was adopted as it produces the optimal performance without overfitting the data.

Mel-Cepstral Distortion (MCD) presented in equation 8 was used for objective evaluation of the conversion success by measuring the spectral distance between the target speech and converted speech.

$$MCD = \frac{10}{Ln(10)} \sqrt{2 \times \sum_{i=1}^{24} (mc_i^t - mc_i^p)^2} \qquad (8)$$

## IV. RESULTS AND DISCUSSION

The overall success of the algorithm for parallel and non-parallel utterances were obtained by calculating the percentage spectral increase or decrease between source - converted MCD and target - converted MCD. This is shown in Fig. 3 and 4 respectively. For Parallel conversion, both source and target speakers uttered the same word while for non-parallel conversion, different words were uttered. The result obtained from non-parallel conversion was slightly better than the one from parallel utterance which was unexpected. To investigate this, the spectral distance between some of the vowels and consonants present in the conversion was measured and the result was presented in Fig. 5. The way the speakers pronounced the vowel sounds were found to have more influence on the result than the consonants. Emphasis in voice conversion must be placed on the articulation of vowel sounds in the word particularly in situations where only a few training samples can be used. This contribution is significant as most voice conversion algorithms such as the ones developed by [28]-[30] never studied this because the training data used were, studio recorded, obtained from trained vocalist and recorded by professional sound directors.

Fig. 5 also revealed the weight of the contribution by the selected vowels with vowel sound "a" being the most dominant. In a noisy environment, it can also be seen that unvoiced consonants like "s", "t" and "f" used in this study degrade the voice conversion success by 55.14 percent relative to the voiced sounds

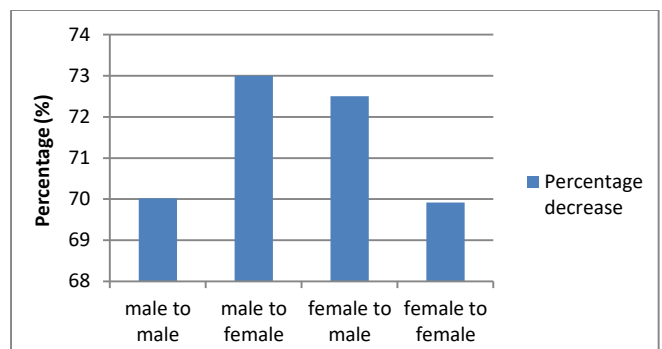
Fig. 3. Conversion Success Rate For Parallel Utterances

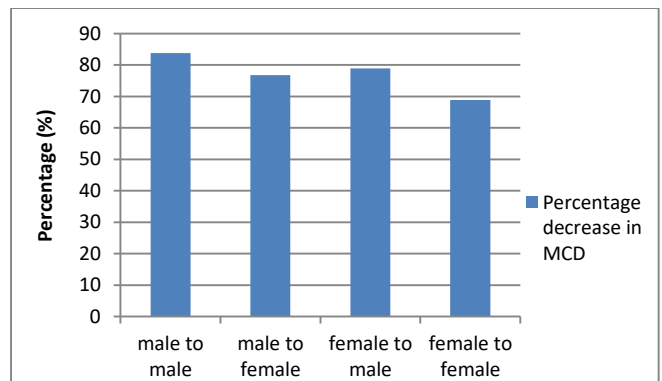
Fig. 4. Conversion Success Rate for Non Parallel Utterances

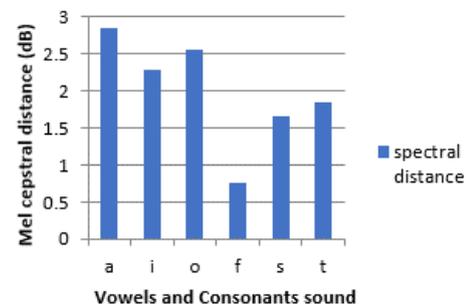
Fig. 5. Average distance between source and target vowels and consonant sounds

Fig. 6 underscores the importance of prosody in voice conversion. The result is more noticeable in conversion where a female is the target. This is significant because females have higher pitches. Hence, conversion situation where the target is a female yielded very poor result without prosody conversion.

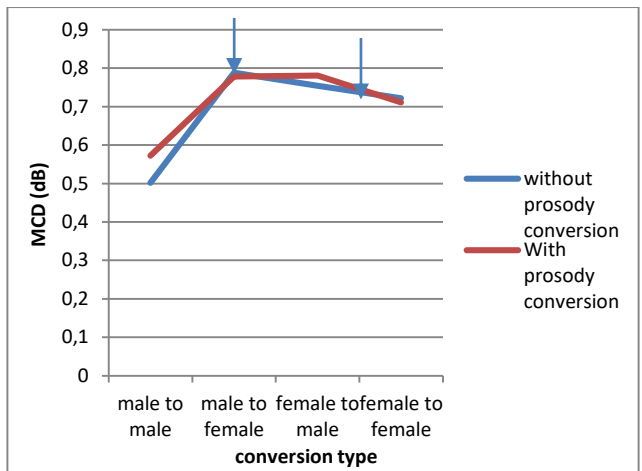

Fig. 6. Impact of prosody in voice conversion

## V. CONCLUSION

This work revealed the contribution of vowels and prosody in neural network based voice conversion model trained with noisy data. Most popular voice conversion models in literature use carefully selected training data obtained from trained voice experts and so the contribution of the speaker's prosody and diction particularly in articulating the voiced and voiceless sounds were overlooked. This study revealed that for voice conversion models trained on samples recorded in uncontrolled noisy environment, the unvoiced sounds which are majorly consonant have more prominent roles to play. The study also revealed that in this kind of scenario, prosody conversion is even more important for male to female cross gender conversion. Further work would involve the study of the percentage contribution of all know voiced and unvoiced sound in the English language as these will be useful in situations where only very few training data are available or allowable for training purposes.